\documentclass{pasj01}
\twocolumn
\Received{$\langle$reception date$\rangle$}
\Accepted{$\langle$acception date$\rangle$}
\Published{$\langle$publication date$\rangle$}
\SetRunningHead{Astronomical Society of Japan}{A Cluster of Galaxies
behind the Galactic Plane Observed with Suzaku and Chandra}

\begin{document}

\title{Suzaku and Chandra Observations of CIZA J1700.8$-$3144, a Cluster
of Galaxies in the Zone of Avoidance}
\author{Hideyuki \textsc{Mori},\altaffilmark{1,2}
Yoshitomo \textsc{Maeda},\altaffilmark{3}
Yoshihiro \textsc{Ueda},\altaffilmark{4}
Kazuhiro \textsc{Nakazawa},\altaffilmark{5}
Yuzuru \textsc{Tawara},\altaffilmark{6}}
\altaffiltext{1}{CRESST and X-ray Astrophysics Laboratory, NASA Goddard
Space Flight Center, Greenbelt, MD 20771, USA}
\altaffiltext{2}{Department of Physics, University of Maryland,
Baltimore County, 1000 Hilltop Circle, Baltimore, MD 21250, USA}
\altaffiltext{3}{Department of Space Astronomy and Astrophysics, Institute of
Space and Astronautical Science (ISAS), Japan Aerospace Exploration
Agency (JAXA), 3-1-1, Yoshinodai, Chuo-ku, Sagamihara, 252-5210}
\altaffiltext{4}{Department of Astronomy, Graduate School of Science,
Kyoto University, Sakyo-ku, Kyoto, 606-8502}
\altaffiltext{5}{Department of Physics, School of Science, The
University of Tokyo, 7-3-1, Hongo, Bunkyo-ku, Tokyo, 113-0033}
\altaffiltext{6}{Division of Particle and Astrophysical Science,
Graduate School of Science, Nagoya University, Furo-cho, Chikusa-ku,
Nagoya, 464-8602}
\email{hideyuki.mori@nasa.gov}

\KeyWords{X-rays: galaxies: clusters${}_1$ --- X-rays: individual (1RXS
J170047.8$-$314442)${}_2$ --- X-rays: individual (CIZA
J1700.8$-$3144)${}_3$}

\maketitle

\begin{abstract}
We present the Chandra and Suzaku observations of 1RXS
 J170047.8$-$314442, located towards the Galactic bulge, to reveal a
 wide-band ($0.3$--$10$~keV) X-ray morphology and spectrum of this
 source.  With the Chandra observation, no point source was found at the
 position of 1RXS J170047.8$-$314442.  Alternatively, we revealed the
 presence of diffuse X-ray emission by the wide-band X-ray image
 obtained from the Suzaku XIS.  Although the X-ray emission had a nearly
 circular shape with a spatial extent of $\sim \timeform{3.5'}$, the
 surface brightness profile was not axisymmetric; a bright spot-like
 emission was found at $\sim \timeform{1'}$ away in the north-western
 direction from the center.  The radial profile of the surface
 brightness, except for this spot-like emission, was reproduced with a
 single $\beta$-model; $\beta$ and the core radius were found to be
 $1.02$ and $\timeform{1.51'}$, respectively.  The X-ray spectrum of the
 diffuse emission showed an emission line at $\sim 6$~keV, indicating an
 origin of a thermal plasma.  The spectrum was well explained with an
 absorbed optically-thin thermal plasma model with a temperature of
 $6.2$~keV and a redshift parameter of $z = 0.14 \pm 0.01$.  Hence, the
 X-ray emission was considered to arise from the hot gas associated with
 a cluster of galaxies.  Our spectroscopic result confirmed the optical
 identification of 1RXS J170047.8$-$314442 by
 \citet{2007ApJ...662..224K}: CIZA J1700.8$-$3144, a member of the
 cluster catalogue in the Zone of Avoidance.  The estimated bolometric
 X-ray luminosity of $5.9 \times 10^{44}$~erg~s$^{-1}$ was among the
 lowest with this temperature, suggesting that this cluster is far from
 relaxed.
\end{abstract}

\section{Introduction}
\label{section:intro}
Population studies of the Galactic components are a key issue to
understand the dynamical formation of the Galaxy.  X-ray sources
containing compact objects, such as neutron stars and black holes, are
useful as a tracer of the primordial high-mass ($M > 8M_{\odot}$)
population.  Thus, focusing on the Galactic bulge
\citep{1999fgb..conf.....C}, we first constructed a flux-limited
($\gtrsim 10^{-12}$~erg~s$^{-1}$~cm$^{-2}$) sample of the X-ray sources
towards this region \citep{2006ESASP.604..459M} extracted from the ROSAT
All-Sky Survey Bright Source Catalogue (RBSC;
\cite{1999A&A...349..389V}).  However, about half ($\sim 40$) of this
bulge sample has been still unidentified.  To improve the completeness
of the sample, our team has been working on the identification of the
Galactic bulge sources with X-ray spectroscopic observations above
$2$~keV (e.g., \cite{2012PASJ...64..112M}).

In this on-going project, we found that one of the unidentified sources
was a cluster of galaxies at $z = 0.13$ \citep{2013PASJ...65..102M}.
Our result implies that clusters of galaxies may occupy a non-negligible
fraction of the bright X-ray sources towards the Galactic bulge.
Because of low Galactic latitudes, moderate X-ray absorption as well as
severe optical extinction hampers detailed spectroscopic studies of the
clusters in the bulge region.

In this paper, we report the results of the Suzaku and Chandra
observations of 1RXS J170047.8$-$314442.  This source was unidentified
when we started our project to clarify the X-ray source population in
the Galactic bulge.  Later, \citet{2007ApJ...662..224K} identified it
with CIZA J1700.8$-$3144, a cluster of galaxies in the CIZA catalogue.
The authors picked up cluster candidates from the RBSC, and then carried
out the $R$-band optical imaging survey to detect their member galaxies.
From the spectroscopic observation to three member galaxies,
\citet{2007ApJ...662..224K} also reported the redshift parameter ($z =
0.134$) of CIZA J1700.8$-$3144.  Because of the limited sensitivity of
the ROSAT PSPC ($0.1$--$2.4$~keV), however, spatial and spectral
information on X-ray emission above $2$~keV associated with this cluster
of galaxies is still unclear.

The paper is organized as follows.  Brief information on the
observations and the data used in the analysis is mentioned in
section~\ref{section:observation}.  We describe the image and spectral
analysis in section~\ref{section:analysis}.  The X-ray properties of the
observed diffuse emission are discussed in
section~\ref{section:discussion}.  Then, a summary of the observations
is given in section~\ref{section:summary}.  Throughout the paper, the
cosmological parameters are assumed to be $H_{0} =
70$~km~s$^{-1}$~Mpc$^{-1}$ (Hubble constant), $\Omega _{\rm M} = 0.27$
(density parameter of matter), and $\Omega _{\lambda} = 0.73$ (density
parameter of dark energy).  Errors represent the 90\% confidence limits
unless otherwise stated.

\section{Observation and data reduction}
\label{section:observation}
\begin{table*}[bhtp]
 \tbl{Observation log.}{%
 \begin{tabular}{lcccc}
 \hline
        & Obs. ID & Start time (UT) & End time (UT) &
 Exposure\footnotemark[$*$] (ks)\\
 \hline
Chandra & 12939\footnotemark[$\dagger$] & 2012/04/20 05:06:32 & 2012/04/20 06:35:26 & 3.69 (ACIS-S3) \\
Suzaku  & 407027010 & 2013/02/18 11:50:13 & 2013/02/18 18:00:11 & 8.93 (XIS), 7.65 (HXD) \\
  \hline
 \end{tabular}}
 \label{table:observation}
\begin{tabnote}
 \footnotemark[$*$] Effective exposure of the screened data. The HXD exposure is
 dead-time corrected.
 \par\noindent
 \footnotemark[$\dagger$] Sequence number is 900971.
\end{tabnote}
\end{table*}

We observed 1RXS J170047.8$-$314442 with Chandra
\citep{2000SPIE.4012....2W} on April 20, 2012.  To handle the expected
source flux, here we operated only the S3 chip of the Advanced CCD
Imaging Spectrometer (ACIS; \cite{1997ITED...44.1633B}) in the very
faint format.  The $128$-pixel subarray model was also applied to avoid
the pile-up effect.  No grating module was employed.  The log of the
Chandra observation is given in table~\ref{table:observation}.  The
level 2 event files generated with the standard data processing were
used in the following analysis.  Scientific products were created with
the Chandra Interactive Analysis of Observations (CIAO, version 4.5:
\cite{2006SPIE.6270E..60F})\footnote{http://cxc.harvard.edu/ciao/}.

The 1RXS J170047.8$-$314442 observation with Suzaku
\citep{2007PASJ...59S...1M} was performed to obtain the X-ray spectrum
above $2$~keV on February 18, 2013.  We show the observation log in
table~\ref{table:observation}.  Suzaku enables us to carry out imaging
spectroscopy in the $0.2$--$10$~keV band with the combination of the
X-ray CCD cameras (XIS; \cite{2007PASJ...59S..23K}) and the X-ray
Telescopes (XRT; \cite{2007PASJ...59S...9S}).  Owing to the XIS2
failure, two front-illuminated (FI) CCDs (XIS0 and XIS3) and one
back-illuminated (BI) CCD (XIS1) were available.  The XIS was operated
with the normal clocking mode without any window and burst options.  The
Spaced-row Charge Injection (SCI; \cite{2009PASJ...61S...9U}) was
applied to mitigate the charge transfer inefficiency.  Suzaku also
possesses a non-imaging hard X-ray detector (HXD;
\cite{2007PASJ...59S..35T}, \cite{2007PASJ...59S..53K}) that consists of
PIN Si diodes and well-type GSO/BGO scintillators to cover an energy
range of $10$--$600$~keV.

We analyzed cleaned event data that had already been pre-processed with
the Suzaku team; the processing version is 2.8.20.35.  While the
exposure time of the XIS was $8.9$~ks, the dead-time corrected one of
the HXD was $7.7$~ks.  We used the HEASOFT version 6.15 and the latest
calibration database to create scientific products.  We note that the
HXD count rate in the $18$--$40$~keV band was $0.160 \pm
0.005$~cts~s$^{-1}$.  The count rate of Non X-ray Background (NXB) was
derived to be $0.153 \pm 0.001$~cts~s$^{-1}$ from
\texttt{ae407027010\_hxd\_pinbgd.evt}, the ``tuned'' background file
provided from the HXD team.  Since the contribution of the Cosmic X-ray
Background (CXB) was estimated to be $0.01$~cts~s$^{-1}$, no significant
signal was detected from the HXD.  Therefore, we focus on the XIS data
in the following analysis.

\section{Analysis}
\label{section:analysis}
\subsection{X-ray image}
\label{subsection:image}

\begin{figure*}[htbp]
 \begin{center}
  \includegraphics[width=80mm]{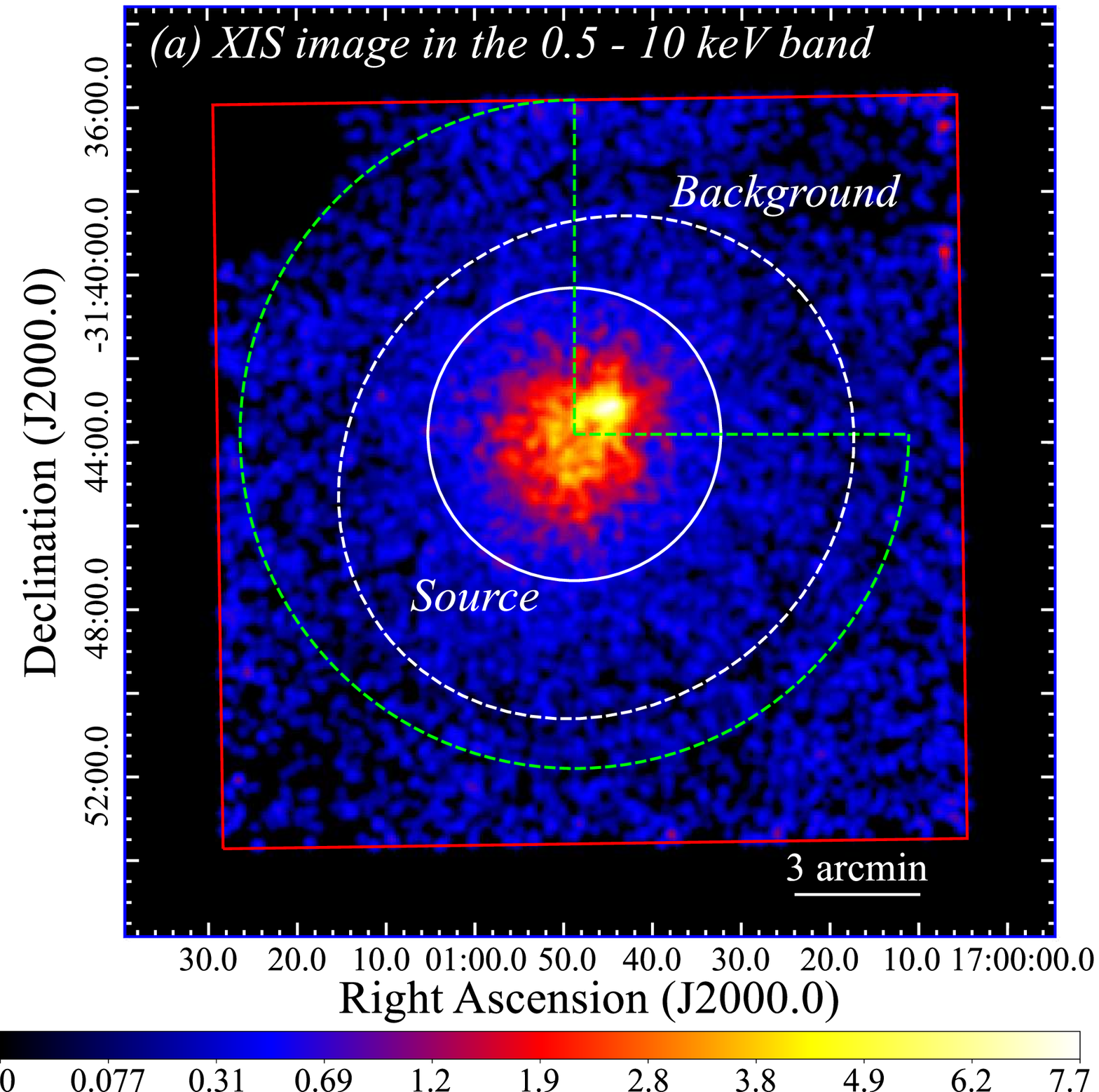}
  \includegraphics[width=80mm]{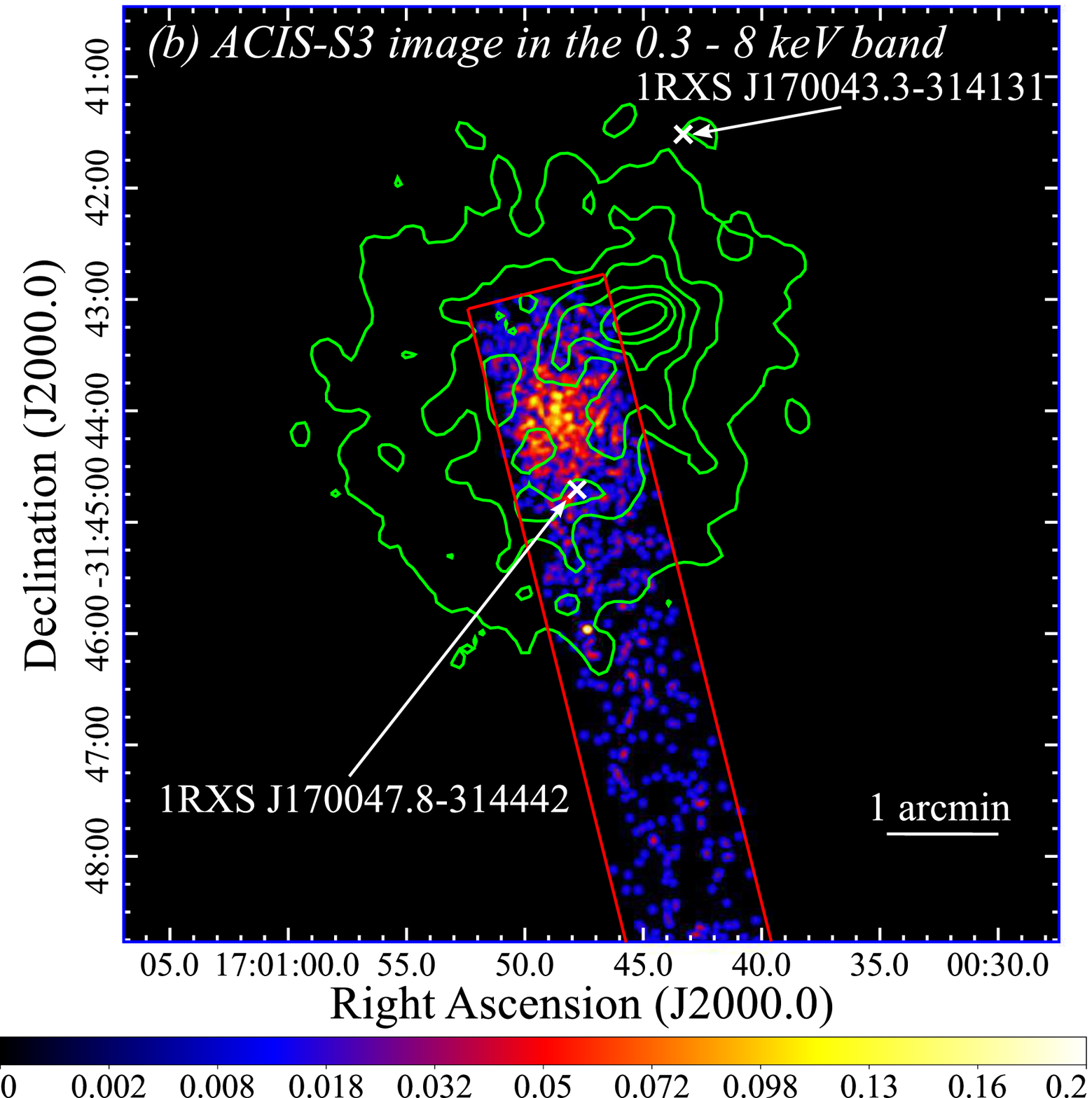}
 \end{center}
\caption{(a) XIS image of 1RXS J170047.8$-$314442 in the $0.5$--$10$~keV
 band. A $\timeform{4.1''} \times \timeform{4.1''}$ binning, the
 smoothing with a Gaussian function of $\sigma = \timeform{12.5''}$, and
 the vignetting correction were applied to the XIS image.  The source
 and background-extraction regions are indicated with a white solid
 circle and its surrounding dashed ellipse, respectively.  The radial
 profile of the X-ray surface brightness (see
 figure~\ref{fig:radial_profile}) was obtained from the green dashed
 sector.  (b) ACIS-S3 image in the $0.3$--$8$~keV band.  The image was
 smoothed with a Gaussian function of $\sigma = \timeform{3''}$.  The
 positions of 1RXS J170047.8$-$314442 and 1RXS J170043.3$-$314131,
 another unidentified X-ray source, listed in the RBSC are represented
 with white crosses.  Green contours show the surface brightness of the
 XIS image.  The contour levels are linearly spaced from $6.6$ to $1.1$
 cts per $\timeform{4.1''} \times \timeform{4.1''}$.  Red squares in
 both panels represent the field of views of the (a) XIS and (b) ACIS-S3
 operated with the $128$-pixel subarray mode.  The color scales at the
 bottom are in unit of (a) photons per $\timeform{4.1''} \times
 \timeform{4.1''}$ and (b) photons per $\timeform{0.5''} \times
 \timeform{0.5''}$.}  \label{fig:Xray_images}
\end{figure*}

The vignetting-corrected XIS image of 1RXS J170047.8$-$314442 in the
$0.5$--$10$~keV band is shown in figure~\ref{fig:Xray_images}a.  We
found diffuse X-ray emission at the center of the XIS field of view,
positionally coincident with 1RXS J170047.8$-$314442.  Thus, we consider
this diffuse emission as 1RXS J170047.8$-$314442 hereafter.  Although
the emission had a nearly circular shape, its surface brightness profile
was not axisymmetric.  The position of the peak brightness was $(RA,
Dec)_{\rm J2000.0}= (\timeform{17h00m44.9s}, \timeform{-31D43'09''})$,
located $\sim \timeform{1'}$ away in the north-western direction from
the center of the X-ray emission.

We also show the ACIS-S3 image in the $0.3$--$8$~keV band in
figure~\ref{fig:Xray_images}b.  No point source was found at the
position of 1RXS J170047.8$-$314442 (white cross in
figure~\ref{fig:Xray_images}b).  Alternatively, we detected extended
emission similar to that in the XIS image.  The brightness peak was
unfortunately located outside of the ACIS-S3 field of view.  However,
the surface brightness around the peak was not so high in the ACIS-S3
image.  Therefore, the size of the north-western spot-like emission was
estimated to be at most $\timeform{30''}$.

Coming back to the XIS data, we indicate regions to extract the source
and background spectra with a white solid circle and a white dashed
ellipse, respectively, in figure~\ref{fig:Xray_images}a.  We selected
the source-extraction region with a radius of $\timeform{3.5'}$ so that
the surface brightness at the boundary reduces to $\sim 5$\% of the peak
brightness.  The center of the circle was chosen to be $(RA, Dec)_{\rm
J2000.0}= (\timeform{17h00m48.8s}, \timeform{-31D43'49''})$.  The major
and minor axes of the surrounding background region were set to be
parallel to the Galactic latitude and longitude, respectively, by taking
into consideration the spatial distribution of the Galactic Ridge X-ray
Emission (GRXE; \cite{1986PASJ...38..121K}).  The outer radii of the
background region were $\timeform{6.43'}$ (major axis) and
$\timeform{5.73'}$ (minor axis).  The redistribution matrix file (RMF)
and ancillary response file (ARF) were created with \texttt{xisrmfgen}
and \texttt{xissimarfgen} \citep{2007PASJ...59S.113I}, respectively.  We
note that the background-subtracted count rates in the $0.5$--$10$~keV
band were $0.26 \pm 0.01$ (XIS0), $0.30 \pm 0.01$ (XIS1), and $0.24 \pm
0.01$ (XIS3) cts~s$^{-1}$.  Since the light curves did not exhibit any
significant variability, we analyzed the summed data during the whole
observation.

\begin{figure}[htbp]
 \begin{center}
  \includegraphics[width=80mm]{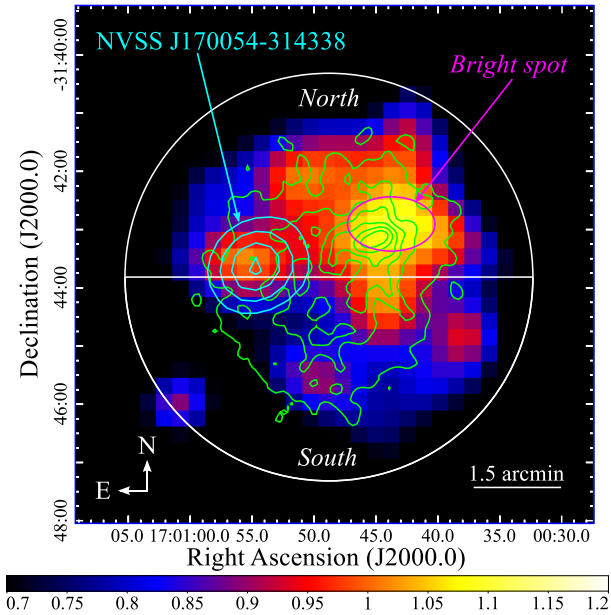}
 \end{center}
 \caption{Color map of the hardness ratio defined as $H/L$, where $H$
 and $L$ represent X-ray photons within $\timeform{17''} \times
 \timeform{17''}$ in the $2$--$10$~keV and $0.5$--$2$~keV bands,
 respectively.  Green contours are the same as those shown in
 figure~\ref{fig:Xray_images}b.  A magenta ellipse indicates the region
 to extract the ``Bright spot'' spectra.  The definition of the
 ``South'' and ``North'' regions is described in the text.  Cyan
 contours display the intensity at $1.4$~GHz from $8$~mJy to $2$~mJy
 obtained from the NRAO VLA Sky Survey (NVSS).  In the NVSS catalogue,
 this radio source is listed as NVSS J170054$-$314348.}
 \label{fig:hr_map}
\end{figure}

To examine spectral properties under poor photon statistics, we further
made a color map of a hardness ratio (HR) given as $H/L$, where $H$ and
$L$ represent the photon counts in the $2$--$10$~keV and $0.5$--$2$~keV
bands, respectively.  The NXB-subtracted images in the soft and hard
X-ray bands were binned with $\timeform{17''} \times \timeform{17''}$.
The HR map is shown in figure~\ref{fig:hr_map}.  The HRs in the X-ray
bright north-western region of 1RXS J170047.8$-$314442 tend to be higher
than those of the south-eastern and outskirt regions.  Thus, we divided
the source-extraction region into three regions, designated as ``Bright
spot'', ``North'', and ``South'', for the following spectral analysis.
Meanwhile, the source-extraction region was defined explicitly to be
``whole source'' region hereafter.  The ``Bright spot'' region was
selected so that the HR was larger than $1.05$.  This region was an
ellipse centered on $(RA, Dec)_{\rm J2000.0}= (\timeform{17h00m43.8s},
\timeform{-31D42'54''})$.  The radii of the major and minor axes of the
``Bright spot'' region were $\timeform{45''}$ and $\timeform{28''}$,
respectively.  While the ``South'' region was defined as the southern
semicircle of the source-extraction region, the ``North'' region was set
to be the northern one excluding the ``Bright spot'' region.

\begin{figure}[htbp]
 \begin{center}
  \includegraphics[width=80mm]{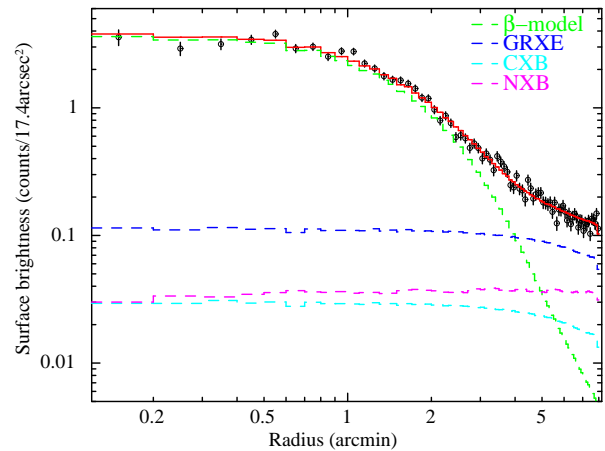}
 \end{center}
 \caption{Radial profile of the surface brightness of 1RXS
 J170047.8$-$314442 in the $0.5$--$10$~keV band (open circles).  A
 vertical bar on each data point indicates a $1 \sigma$ statistical
 error.  The red solid line represents the best-fit combined model that
 consists of the $\beta$-model with $(\beta, r_{\rm c}) = (1.02,
 \timeform{1.51'})$ (green), GRXE (blue), CXB (cyan), and NXB (magenta
 dashed line) components.}
 \label{fig:radial_profile}
\end{figure}

In order to calculate the ARF accurately, we need a 2-dimensional
distribution of the X-ray intensity of 1RXS J170047.8$-$314442.  Since
the observed X-ray emission was not axisymmetric, we first made a radial
profile of the surface brightness from a sector shown in
figure~\ref{fig:Xray_images}a, where the north-western bright spot was
removed.  The resultant radial profile is shown in
figure~\ref{fig:radial_profile}.  We fitted the profile with a single
$\beta$-model \citep{1976A&A....49..137C}, taking into account the point
spread function (PSF) of the XRT and the contribution from the
underlying diffuse emission: NXB, GRXE, and CXB.  The fitting
prescription we used here was the same as that described in
\citet{2013PASJ...65..102M}.  The combined model of $(\beta, r_{\rm c})
= (1.02, \timeform{1.51'})$ yielded a marginally acceptable fit with
$\chi ^{2} = 70/48$~d.o.f..  The $68$\% confidence limits were $\beta =
1.02^{+0.10}_{-0.08}$ and $r_{\rm c} =
\timeform{1.51'}^{\timeform{+0.16'}}_{\timeform{-0.15'}}$.

Since the angular resolutions of the Suzaku XRTs are limited to be $\sim
\timeform{2'}$ \citep{2007PASJ...59S...9S}, the spatial structure of the
north-western ``Bright spot'' region cannot be fully resolved.  Thus,
for simplicity of the ARF calculation, we modelled this emission as a
single point source.  After subtracting the simulated images of the
GRXE, CXB, NXB, and $\beta$-model components from the raw XIS image, we
determined the source position to be $(RA, Dec)_{\rm J2000.0} =
(\timeform{17h00m44.6s}, \timeform{-31D43'09''})$.  The normalization of
the point source was set so as to reproduce the ratio of the observed
X-ray counts between the point source and $\beta$-model components.  We
made hereafter some ARF files using a sky image that consists of the
$\beta$-model with $(\beta, r_{\rm c}) = (1.02, \timeform{1.51'})$ and
this point source.  Hence, X-ray spectra represent a weighted average of
the source spectrum over the extraction region we chose.

\subsection{X-ray spectrum}
\label{subsection:spectrum}
\begin{figure}[htbp]
 \begin{center}
  \includegraphics[width=80mm]{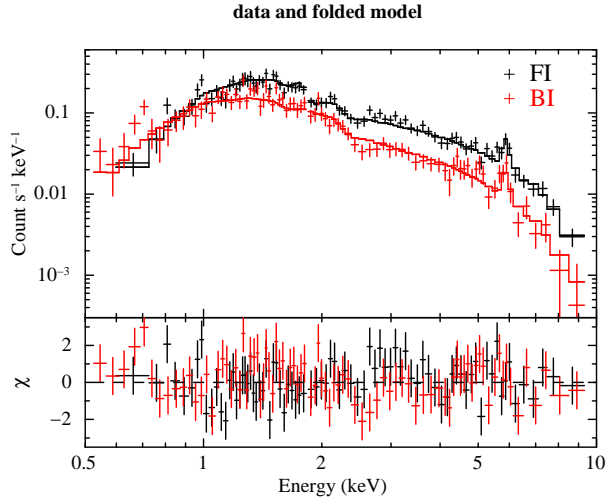}
 \end{center}
 \caption{X-ray spectra of 1RXS J170047.8$-$314442 obtained with the FI
 (black pluses) and BI (red pluses) CCDs.  The best-fit optically-thin
 thermal plasma model is indicated with the black and red solid lines.}
 \label{fig:1RXSJ170047_spectra}
\end{figure}

The X-ray spectra of the whole source region obtained with the FI and BI
CCDs are shown in figure~\ref{fig:1RXSJ170047_spectra}.  The XIS0 and
XIS3 spectra were averaged to increase the photon statistics.  In the
spectral fit, we excluded spectral bins around the Si-K edge
($1.82$--$1.84$~keV) because of the uncertainty of the quantum
efficiency.  The presence of the clear emission line at $5.9$~keV
indicates that the X-ray source is not an object in the Galaxy.  In
addition, the spatially extended X-ray emission combined with its convex
shape suggests that the source has a thermal origin.  Then, we fitted
the spectra with an optically-thin thermal plasma model in collisional
ionization equilibrium (\texttt{apec} in \texttt{XSPEC}) affected by
photoelectric absorption (\texttt{phabs} in \texttt{XSPEC};
\cite{1992ApJ...400..699B}).  Here, we adopted the solar abundance given
by \citet{1989GeCoA..53..197A}.  We also set the redshift parameter
free.  The absorbed thin-thermal plasma model gave a good fit with $\chi
^{2} = 177/161$~d.o.f. and null hypothesis probability of $19$\%.  The
best-fit parameters are summarized in table~\ref{table:1RXSJ170047_fit}.
The absorption column density, plasma temperature, metal abundance, and
redshift parameter were $N_{\rm H} = (2.4 \pm 0.4) \times
10^{21}$~cm$^{-2}$, $kT = 6.2 \pm 0.7$~keV, $Z/Z_{\odot} = 0.26 \pm
0.08$, and $z = 0.14 \pm 0.01$, respectively.  The X-ray flux in the
$0.5$--$10$~keV band was $6.0 \times 10^{-12}$~erg~s$^{-1}$~cm$^{-2}$.

\begin{table*}
\tbl{Best-fit parameters of the 1RXS J170047.8$-$314442
spectra\footnotemark[$*$].}{%
\begin{tabular}{lllll}
\hline
Parameters  & \multicolumn{4}{c}{Model : \texttt{apec}\footnotemark[$\dagger$]} \\
            & Whole source & South & North & Bright spot \\
\hline
$N_{\rm H}$ ($10^{21}$ cm$^{-2}$) & $2.4 \pm 0.4$ & $2.5 \pm 0.6$
 & $2.4^{+0.7}_{-0.5}$ & $2.6^{+1.6}_{-1.2}$\\
$kT$ (keV)                & $6.2 \pm 0.7$ & $5.0^{+0.7}_{-0.6}$
 & $6.9^{+1.6}_{-0.9}$ & $10^{+9}_{-4}$\\
$Z/Z_{\odot}$\footnotemark[$\ddagger$] & $0.26 \pm 0.08$ &
 $0.35^{+0.13}_{-0.12}$ & $0.24^{+0.14}_{-0.13}$ & $0.26$ (fixed)\\
Redshift $z$                      & $0.14 \pm 0.01$ & $0.137 \pm 0.008$
 & $0.14^{+0.03}_{-0.01}$ & $0.14$ (fixed) \\
$EM$\footnotemark[$\S$]           & $\bigl ( 6.0^{+0.4}_{-0.3} \bigr ) \times 10^{-3}$
     & $\bigl ( 6.1^{+0.5}_{-0.6} \bigr ) \times 10^{-3}$
     & $\bigl ( 5.6^{+0.5}_{-0.2} \bigr ) \times 10^{-3}$
     & $\bigl ( 5.9^{+1.1}_{-0.6} \bigr ) \times 10^{-3}$\\
\hline
$\chi ^{2}$/d.o.f.                & $177/161 = 1.1$ & $139/137 = 1.0$ &
 $126/126 = 1.0$ & $48/33 = 1.5$ \\
Flux\footnotemark[$\l$] ($10^{-12}$~erg~s$^{-1}$~cm$^{-2}$) & $6.0$ & $5.7$
 & $5.9$ & $7.0$ \\
\hline
\end{tabular}}
\label{table:1RXSJ170047_fit}
\begin{tabnote}
 \footnotemark[$*$] Superscript and subscript figures represent the
upper and lower limits of the 90\% confidence interval, respectively.
 \par\noindent
 \footnotemark[$\dagger$] Model components defined in \texttt{XSPEC}.
 \par\noindent
 \footnotemark[$\ddagger$] Elemental abundances relative to solar \citep{1989GeCoA..53..197A}.
 \par\noindent
 \footnotemark[$\S$] In unit of $10^{-14}/(4 \pi D^{2}) \int n_{\rm e}
n_{\rm H} dV$~cm$^{-5}$. Here $V$ and $D$ are the volume and distance
to the plasma, respectively.
 \par\noindent
 \footnotemark[$\l$] X-ray fluxes in the $0.5$--$10$~keV band.
\end{tabnote}
\end{table*}

\begin{figure}[htbp]
 \begin{center}
  \includegraphics[width=80mm]{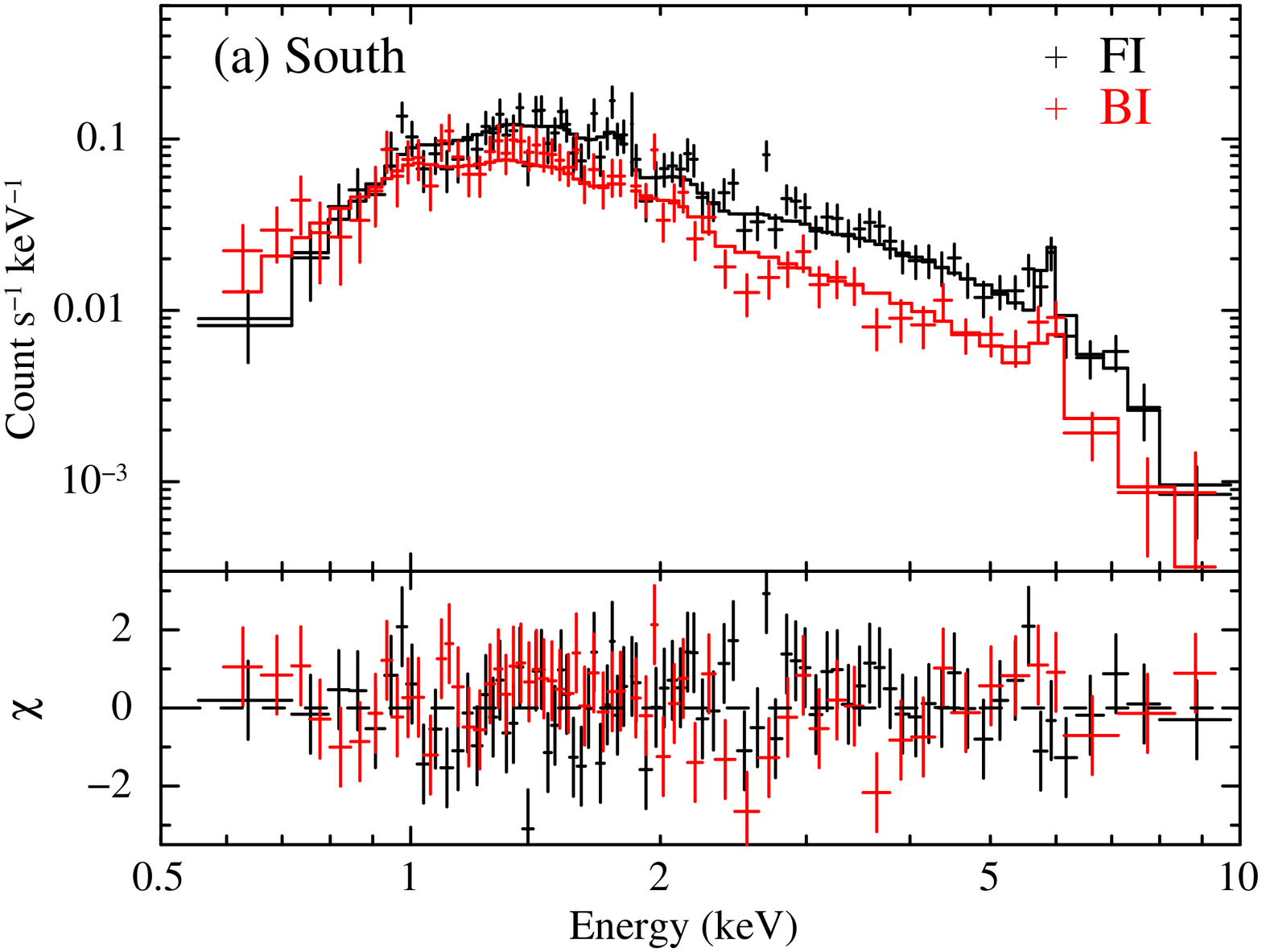} \\
  \includegraphics[width=80mm]{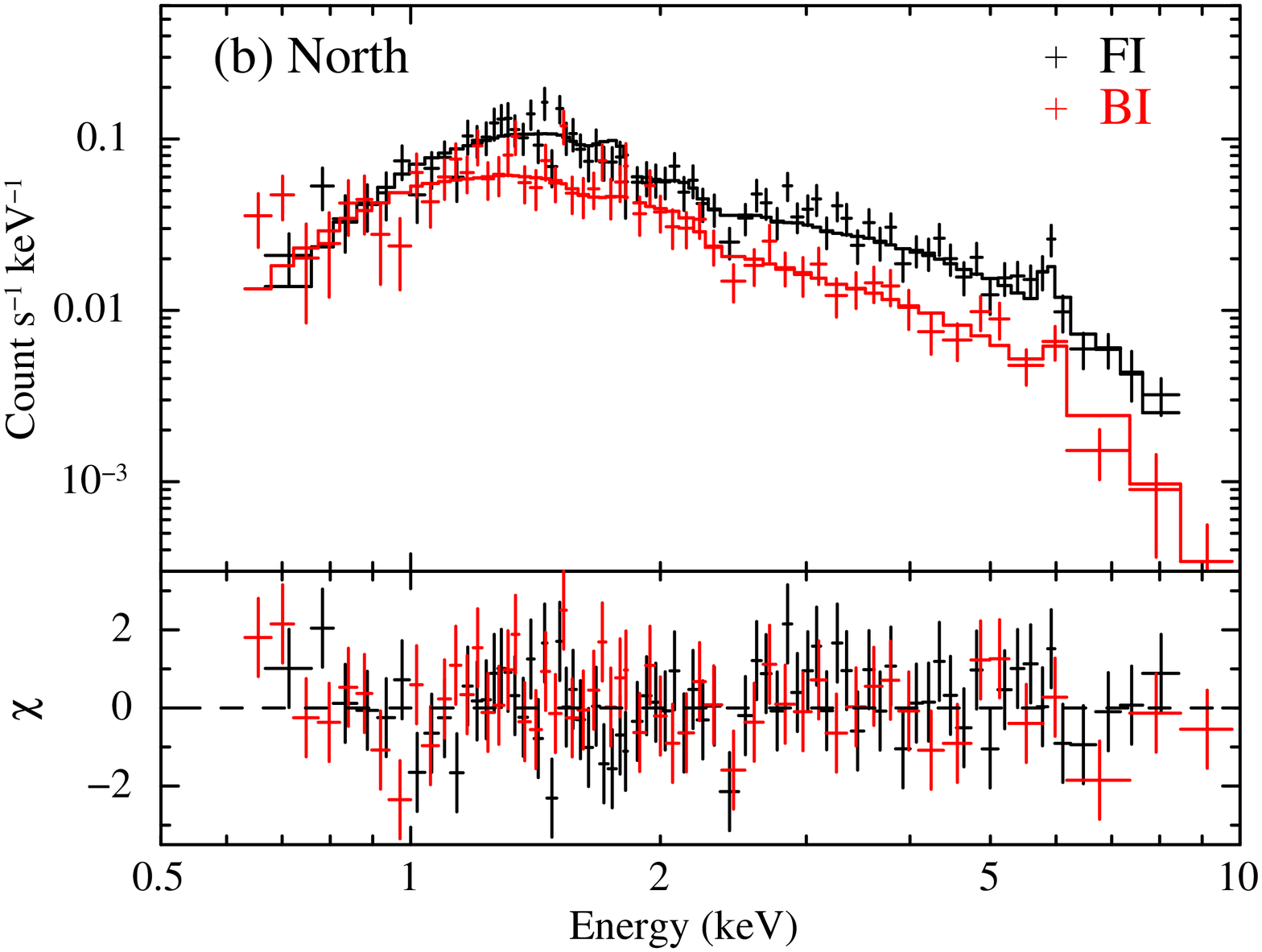} \\
  \includegraphics[width=80mm]{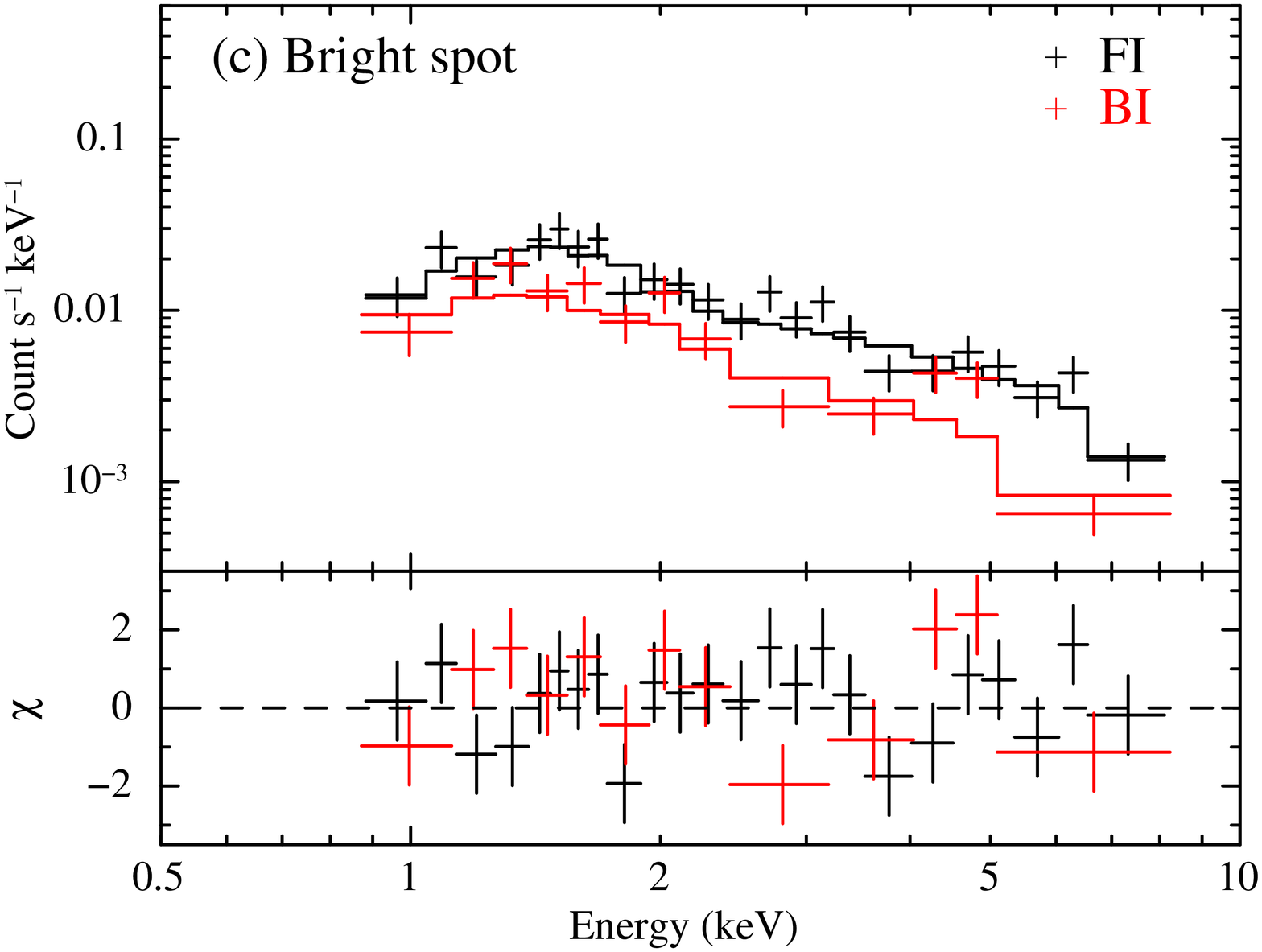}
 \end{center}
 \caption{X-ray spectra of 1RXS J170047.8$-$314442 extracted from the
 (a) ``South'', (b) ``North'', and (c) ``Bright spot'' regions.  Marks
 and colors are the same as those in
 figure~\ref{fig:1RXSJ170047_spectra}.}
 \label{fig:1RXSJ170047_detailed_spectra}
\end{figure}

Next, we fitted the ``South'', ``North'', and ``Bright spot'' spectra
with an absorbed optically-thin thermal plasma model, as is shown in
figure~\ref{fig:1RXSJ170047_detailed_spectra}.  The background spectrum
for these spectra was the same as that extracted from the background
region.  While the ``Bright spot'' spectra were grouped to contain at
least $20$~photons in each spectral bin, the ``South'' and ``North''
spectra were binned with $30$~photons for the FI and $25$~photons for
the BI.  For the ``South'' and ``North'' spectra, we obtained acceptable
fits with $\chi ^{2} = 139/137$~d.o.f. (South) and $\chi ^{2} =
126/126$~d.o.f. (North).  The plasma temperature in the ``South'' region
was found to be significantly lower than that in the ``North'' region
(see table~\ref{table:1RXSJ170047_fit}) , although its significance was
at most $2.9 \sigma$ level.

Since the emission feature at $\sim 6$~keV was obscure in the ``Bright
spot'' spectra, we fixed the redshift and metal abundance to be $0.14$
and $0.26$, respectively.  Although the model yielded a marginally
acceptable fit of $\chi ^{2} = 48/33$~d.o.f., the best-fit temperature
of $kT = 10^{+9}_{-4}$~keV was again larger than that in the ``South''
region ($kT = 5.0^{+0.7}_{-0.6}$~keV) with $1.9 \sigma$ significance.

Furthermore, we tried to fit the ``Bright spot'' spectra with several
different models.  When employing the absorbed thin-thermal plasma
model, in which the redshift and metal abundance were set to be free,
the $\chi ^{2}$ value decreased to be $43$ ($31$~d.o.f.).  The best-fit
parameters did not change, except for the redshift parameter: $z =
0.41^{+0.07}_{-0.08}$.  This larger redshift parameter would be caused
by a line-like structure around $4.5$~keV only shown in the BI spectrum.
However, the $F$-test resulted in a chance probability due to the
statistical fluctuation of $16$\%, and hence we conclude it is not
significant enough.  We note that the spectra were also reproduced with
an absorbed power-law model ($\chi ^{2} = 47/33$~d.o.f.); the best-fit
parameters were $N_{\rm H} = (4.1^{+1.9}_{-1.8}) \times
10^{21}$~cm$^{-2}$ and $\Gamma = 1.8 \pm 0.2$.  Consequently, owing to
the limited photon statistics, such simple models were also marginally
acceptable, but did not give a significantly improved fit.

\section{Discussion}
\label{section:discussion}
We found diffuse X-ray emission above $2$~keV, for the first time, from
the Suzaku and Chandra observations of 1RXS J170047.8$-$314442.  The
improvement of photon statistics with both satellites allows us to
characterize quantitatively X-ray properties on this diffuse emission.
Chandra revealed that there was no point source at the position listed
in the RBSC.  Moreover, at least within the ACIS-S3 field of view, no
point sources were found to contribute to the diffuse X-ray emission.
This diffuse emission had a circular shape with a spatial extent of
$\sim \timeform{3.5'}$, much larger than the ROSAT PSF.  In addition,
the center position of the source was located $\sim \timeform{1'}$ away
in the north direction from the RBSC position.  The source extent ($\sim
\timeform{3.5'}$) might cause the source characterization of 1RXS
J170047.8$-$314442 to be somewhat inaccurate.

We discovered that the X-ray emission of 1RXS J170047.8$-$314442 was not
axisymmetric and that the peak of the X-ray surface brightness was
shifted by $\sim \timeform{1'}$ in the north-western direction from the
center.  The Suzaku XRTs did not enable us to resolve this spot-like
excess emission; the possibility that a point source was superposed by
chance in the line of sight to the diffuse source cannot be ruled out.
The bright spot was located $\sim \timeform{15''}$ away from the Chandra
ACIS-S3 field of view, and then no bright emission related to the spot
was detected.  Thus, the spot size was deduced to be $\lesssim
\timeform{30''}$.  We searched the SIMBAD astronomical database
\footnote{http://simbad.u-strasbg.fr/simbad} for other X-ray sources in
the field of views of the Suzaku and Chandra observations.  Only one
unidentified X-ray source, 1RXS J170043.3$-$314131, was found.  However,
the source position is $\sim \timeform{1.4'}$ away from the bright spot
in the north direction, as is shown in figure~\ref{fig:Xray_images}b.
Assuming that excess X-ray emission was associated with the
circular-shaped extended source, we modelled the non-axisymmetric
surface brightness profile of 1RXS J170047.8$-$314442 with a
$\beta$-model component of $(\beta, r_{\rm c}) = (1.02,
\timeform{1.51'})$ plus a single point source.

The X-ray spectrum of 1RXS J170047.8$-$314442 was found to show an
emission line at $\sim 6$~keV.  The XIS spectra were reproduced well
(null hypothesis probability of $19$\%) with an absorbed thin thermal
plasma model with $N_{\rm H} = (2.4 \pm 0.4) \times 10^{21}$~cm$^{-2}$
and $kT = 6.2 \pm 0.7$~keV.  The averaged H\emissiontype{I} column
density towards the source is $1.8 \times 10^{21}$~cm$^{-2}$
\citep{1990ARA&A..28..215D}, significantly smaller than the fitted
absorption column density.  However, $N_{\rm H}$ is still in a range of
the spatial variation of the H\emissiontype{I} column density.  Hence,
the X-ray absorption is probably attributed to the Galactic interstellar
medium.  Assuming that the emission line was originated from highly
ionized Fe ions (Fe\emissiontype{XXV}), the redshift parameter was
determined to be $z = 0.14$.

The spatial morphology of the X-ray emission and the X-ray spectral
properties strengthen that 1RXS J170047.8$-$314442 is indeed a cluster
of galaxies, identified with CIZA J1700.8$-$3144.  The X-ray determined
redshift of $z = 0.14 \pm 0.01$, consistent with that of CIZA
J1700.8$-$3144 ($z = 0.134$), corresponds to the luminosity distance of
$660$~Mpc.  The angular separation of $\timeform{1''}$ is converted into
$2.5$~kpc.  Therefore, the core radius of $r_{\rm c} = \timeform{1.51'}$
is $220$~kpc.

We estimated the total mass of the cluster, using the best-fit $\beta$
model with $(\beta, r_{\rm c}) = (1.02, \timeform{1.51'})$.  Here the
hydrostatic equilibrium and isothermality of the X-ray emitting gas were
assumed.  The total mass within a given radius is represented with
\begin{equation}
M (< r) = \frac{3 \beta k T_{\rm e} r^{3}}{G \mu m_{\rm H} r^{2}_{\rm
 c}} \frac{1}{1 + (r/r_{\rm c})^2},
\end{equation}
where $k$, $G$, $\mu = 0.6$, and $m_{\rm H}$ represent the Boltzmann
constant, the gravitational constant, the mean molecular weight, and the
mass of the hydrogen, respectively \citep{1980ApJ...241..552F}.  The
total mass within the whole source region was $M (< \timeform{3.5'}) =
(3.1 \pm 0.9) \times 10^{14}$~$M_{\odot}$, where the error represents $1
\sigma$ confidence limit.  We further calculated the overdensity radii
of $r_{500}$ and $r_{200}$, within which the averaged density is $500$
and $200$ times larger than the critical density of the universe ($\rho
_{\rm crit} = 9.2 \times 10^{-30}$~g~cm$^{-3}$).  $r_{500}$ and
$r_{200}$ were evaluated to be $7.0 r_{\rm c} = 1.6$~Mpc and $11 r_{\rm
c} = 2.5$~Mpc, respectively.

To compare our result with scaling laws of clusters, e.g., $L_{\rm
X}$--$T$ relation, in the literature, we evaluated the unabsorbed
bolometric luminosity within $r_{\rm 500}$.  The ARF was re-calculated
for a circular region with a radius of $r_{\rm 500} = \timeform{10.6'}$
centered on $(RA, Dec)_{\rm J2000.0}= (\timeform{17h00m48.8s},
\timeform{-31D43'49''})$, the same as that of the whole source region.
Using the new ARF, the unabsorbed bolometric flux and luminosity were
deduced to be $(1.1 \pm 0.1) \times 10^{-11}$~erg~s$^{-1}$~cm$^{-2}$ and
$L_{\rm X} (r < r_{500}) = (5.9 \pm 0.7) \times 10^{44}$~erg~s$^{-1}$,
respectively.  According to the $L_{\rm X}$--$T$ relation derived from a
cluster sample obtained with the Chandra observations
\citep{2012MNRAS.421.1583M}, this luminosity was by a factor of $\sim 3$
lower than that expected from the X-ray emitting plasma with $kT =
6.2$~keV.  The authors pointed out that unrelaxed systems obeyed the
$L_{\rm X}$--$T$ relation with a steeper slope and a lower normalization
than those of self-similar relaxed ones.  The bolometric luminosity of
$L_{\rm X} \sim 6 \times 10^{44}$~erg~s$^{-1}$ was indeed consistent
with that estimated from the $L_{\rm X}$--$T$ relation of their
unrelaxed cluster sample, by taking into consideration an intrinsic
scatter of $L_{\rm X}$ for this subsample.  Hence, CIZA J1700.8$-$3144
would be a cluster which is not dynamically relaxed yet.

Because of the limited photon statistics, we found no spatial variation
of the plasma temperature at $> 3 \sigma$ confidence level.  Thus, in
order to reveal the plasma dynamics, follow-up X-ray observations with
longer exposures should be required.  Because of the large collecting
area and the good angular resolution of $\sim \timeform{15''}$,
XMM-Newton will be an appropriate observatory for the detailed
spectroscopic study of CIZA J1700.8$-$3144.  Furthermore, it should be
necessary to conduct follow-up optical observations in more detail, in
order to pick up more member galaxies.  The spatial distribution and
kinematics of the member galaxies may be available as a tracer of the
dynamical process.

A deep radio observation will also be encouraged to unveil the dynamics
of this cluster of galaxies.  Radio halos or radio relics indicate the
presence of high-energy electrons that may be originated from shock
acceleration (e.g., \cite{2012A&ARv..20...54F}).  We note that there is
an extended radio source, NVSS J170054$-$314338
\citep{1998AJ....115.1693C}, located near the center of CIZA
J1700.8$-$3144 ($\sim \timeform{1.1'}$ away in the eastern direction,
see cyan contours of figure~\ref{fig:hr_map}).  Nevertheless, the size
of NVSS J170054$-$314338 is $\timeform{31''}$ along the major axis and
$\timeform{25''}$ along the minor axis, much smaller than that of CIZA
J1700.8$-$3144.  Hence, the radio source may be neither radio halo nor
radio relic associated with the cluster of galaxies.

\section{Summary}
\label{section:summary}
We found that no point source was located at the position of 1RXS
J170047.8$-$314442 with the Chandra observation.  Instead, the Suzaku
observation of 1RXS J170047.8$-$314442 revealed the presence of
non-axisymmetric diffuse X-ray emission with a spatial extent of $\sim
\timeform{3.5'}$.  The X-ray emission had a bright spot shifted by $\sim
\timeform{1'}$ in the north-western direction from its center.  Except
for this bright spot-like emission, the surface brightness profile of
1RXS J170047.8$-$314442 can be explained with a single $\beta$-model
with $(\beta, r_{\rm c}) = (1.02, \timeform{1.51'})$.

The X-ray spectrum of 1RXS J170047.8$-$314442 was well reproduced with
an absorbed optically-thin thermal plasma model; the absorption column
density and the plasma temperature were determined to be $(2.4 \pm 0.4)
\times 10^{21}$~cm$^{-2}$ and $6.2 \pm 0.7$~keV, respectively.  Judging
from the absorption column density, the X-ray source should be an
extragalactic object.  Assuming that the emission line at $\sim 6$~keV
is due to Fe\emissiontype{XXV} K$\alpha$ emission, we estimated the
X-ray redshift of $z = 0.14 \pm 0.01$.

The spatial morphology and spectral properties support that the origin
of the X-ray emission is a hot gas associated with a cluster of
galaxies, confirming the optical identification with CIZA J1700.8$-$3144
($z = 0.134$; \cite{2007ApJ...662..224K}).  The unabsorbed bolometric
luminosity of $L_{\rm X} (r < r_{500}) = (5.9 \pm 0.7) \times
10^{44}$~erg~s$^{-1}$, which was consistent with that expected from the
$L_{\rm X}$--$T$ relation of the unrelaxed clusters
\citep{2012MNRAS.421.1583M}, indicates that CIZA J1700.8$-$3144 may be
in an unrelaxed state.

\begin{ack}
First of all, we are deeply grateful to Dr. Harald Ebeling for careful
 and fruitful comments to improve our manuscript.  We would like to
 thank all the Suzaku team members for their support of the observation
 and useful information on the XIS and HXD analyses.  The scientific
 results reported in this paper are also based on the observation made
 by the Chandra X-ray Observatory.
\end{ack}

\end{document}